\documentstyle[aps,epsfig,prd,preprint]{revtex} 
\def \beq{\begin{equation}}         \def \eeq{\end{equation}}
\def \beqa{\begin{eqnarray}}        \def \eeqa{\end{eqnarray}}
\def \bea{\begin{array}}	    \def \eea{\end{array}}

\def\plb#1#2#3{    {\it Phys. Lett. }{\bf B#1} (#2) #3}
\def\prd#1#2#3{    {\it Phys. Rev. }{\bf D#1} (#2) #3}

\def\prl#1#2#3{    {\it Phys. Rev. Lett. }{\bf #1} (#2) #3}
\def\ijm#1#2#3{    {\it Int.J.Mod.Phys.}{\bf A#1}  (#2) #3}  

\def\zpc#1#2#3{    {\it Zeit. f{\"u}r Physik }{\bf C#1} (#2) #3}

\def\rarrow{\rightarrow}  \def\nomb{\nonumber}
\def \pt{\left| {P\over T} \right|}
\def \ptt{\left| {P'\over T'} \right|}

\def \abs#1{\left| #1 \right|}

\begin{document}

\title{ Weak Phase $\gamma$ and Strong Phase $\delta$  from \\ CP 
       Averaged $B\rightarrow \pi\pi\ $ and $\pi K$ Decays }
\author{{ Yue-Liang Wu and Yu-Feng Zhou }\\
  {\small Institute of Theoretical Physics, Chinese Academy of Science, \\ 
Beijing 100080,  China}}
\date{ylwu@itp.ac.cn, zhouyf@itp.ac.cn}
\maketitle
\draft
\begin{abstract}
 Assuming $SU(3)$ symmetry for the strong phases in the four decay modes $B\rarrow 
 \pi^-\pi^+, \pi^0 \pi^+ , \pi^- K^+, \pi^- \bar{K}^0$ and ignoring the relative 
small electroweak penguin effects in those decays, the weak phase $\gamma$ 
and the strong phase $\delta$ can be determined in a model independent way by the 
CP-averaged branching ratios of the four decay modes. It appears that the current
experimental data for $ B\rightarrow \pi\pi$ and $\pi K$ decays prefer a negative value 
of $\cos\gamma\cos\delta $. By combining with the other constraints from $V_{ub}$, 
$B^{0}_{d,s}-\bar{B}^{0}_{d,s}$ mixings and indirect CP-violating parameter $\epsilon_K$ 
within the standard model, two favorable solutions for the phases $\gamma$ and $\delta$ 
are found to lie in the region:
$35^{\circ}\alt\gamma\alt 62^{\circ}$ and $106^{\circ}\alt \delta \alt 180^{\circ}$ or 
$86^{\circ}\alt\gamma\alt 151^{\circ}$ and $0^{\circ}\alt\delta\alt 75^{\circ} $ 
within 1$\sigma$ standard deviation. It is noted that if allowing the standard deviation 
of the data to be more than 1$\sigma$, the two solutions could approach to one solution 
with a much larger region for the phases $\gamma$ and $\delta$. Direct CP asymmetry 
$ a_{\epsilon''}^{(\pi^- K^+)}$ in $B\rarrow \pi^-K^+$ decay can be as large as the present
experimental upper bound. Direct CP asymmetry $ a_{\epsilon''}^{(\pi^+\pi^-)}$
in $B\rarrow \pi^-\pi^+$ decay can reach up to about $40\%$ at 1$\sigma$ level.  
\end{abstract}

\newpage

The study of CP-violation is one of the central topics in the present day 
elementary particle physics. In the standard model (SM), all the CP violating phenomena
arise from a single complex phase of the Cabbibo-Kobayashi-Maskawa(CKM)
matrix elements. If the CKM phase is the only source of CP violation, some unitarity relations such as 
$V_{ud}V^*_{ub}+V_{cd}V^*_{cb}+V_{td}V^*_{tb}=0$ will hold . The unitarity relations can 
be represented geometrically by a set of triangles called unitarity
 triangles . The three angles in the triangle containing $b$ and $d$ quarks are often denoted by 
 $\alpha,\beta$ and $\gamma$ with $\alpha +\beta + \gamma=\pi$ in the SM.
 Thus one of the important issues is to precisely determine those
 angles and their sum. Any deviation of the sum from $\pi$ will be a signal 
 of new physics\cite{newphys}. 
 
 Although angles $\alpha$ and $\beta$ may be well measured via the time dependent 
 measurements of $B\rarrow \psi K_S$ and $B\rarrow \pi\pi$, the determination of
 angle $\gamma$ is a great challenge for both theorists and experimentists. In the recent
 years much work has been done on this issue\cite{wyler91,PW,rosner,fleischer}. As it was first proposed 
 in Ref.\cite{wyler91}, the angle $\gamma$ may be extracted through  six $B\rarrow DK$ 
 decay rates. The difficulty of this method is that it needs tagging of the CP eigenstate
 $D^0_{CP}$ which is rather difficult in the experiment. It may also be extracted from the 
 decay mode $B_s^0 \rightarrow (D_s^- K^+, D_s^+ K^-) \leftarrow\bar{B}_s^0$ in 
 a model-independent way\cite{PW} since one only needs to extract the rephase invariant observables 
 $a_{\epsilon + \epsilon'}$ and $a_{\epsilon'}$ from a time-dependent measurement. Thus the 
 weak phase is simply given by $\sin \gamma = a_{\epsilon + \epsilon'}/
 \sqrt{1 + a_{\epsilon'}^2} $. 
 In the recent years an alternative way of using the CP averaged  
 $B\rarrow \pi^{\pm}K^0, \pi^+\pi^0$ and the CP conjugate 
$B^+\rarrow \pi^0 K^+,\ B^-\rarrow \pi^0 K^-$
 branching ratios has been aroused a great attention\cite{rosner}. However this method 
needs some theoretical input in evaluating the electroweak penguin (EWP) effects. 
At present, limited by the statistics the difference of CP conjugate rates can not be 
definitely established\cite{cleoCP} 
 
  Recently, the CLEO Collaboration has reported the first observation of rare decays
$B\rarrow \pi^-\pi^+$ and $\pi^0\bar{K}^0$ \cite{CLEOold}. The observation of $\pi^0 \bar{K}^0 $
complete the set of measurements on $B\rarrow \pi K$ decays. Other channels of $\pi K$
have also been largely improved. The most recent results reported by CLEO collab. are (in
units of $10^{-6}$)\cite{CLEO},
\beqa
Br(B\rarrow \pi^-\pi^+)&=&4.3^{+1.6}_{-1.4}\pm 0.5 \nomb\\
Br(B\rarrow \pi^0 \pi^+)&=&<12.7 \ \ \ (5.6^{+2.6}_{-2.3}\pm 1.7) \nomb\\
Br(B\rarrow \pi^- K^+)&=&17.2^{+2.5}_{-2.4}\pm 1.2 \nomb\\
Br(B\rarrow \pi^- \bar{K}^0)&=&18.2^{+4.6}_{-4.0}\pm 1.6 \nomb\\
Br(B\rarrow \pi^0 K^+)&=& 11.6^{+3.0+1.4}_{-2.7-1.3} \nomb\\
Br(B\rarrow \pi^0 \bar{K}^0)&=&14.6^{+5.9+2.4}_{-5.1-3.3}
\eeqa

Although only the upper bound of $\pi^+\pi^0$ is given, the CLEO Collab. also
quote a value of $Br(B\rarrow \pi^0 \pi^+)=5.6^{+2.6}_{-2.3}\pm 1.7$. This will be improved by the future
measurements. The relative small value of $\pi^-\pi^+$, the almost equal $K \pi$ rates:
$\pi^- \bar{K}^0 \approx \pi^- K^+$ and large  $\pi^0 \bar{K}^0$ 
seem to be in conflict with the theoretical predictions. However, as it was
pointed out in Ref.\cite{he99prl}, if one takes the weak phase $\gamma$ of the CKM
matrix elements to be larger than $90^{\circ}$ and include the EWP
effects, the situation for $\pi^0K^+$ may be improved greatly, but for $\pi^0\bar{K}^0$ it may 
become worse as the EWP-SP (strong penguin) interference in $\bar{K}^0\pi^0$ decay is likely
to be destructive. Some alternative ways in solving
this puzzle are also proposed, such as the small $|V_{ub}|$ in $B\rarrow \pi^+
\pi^-$\cite{Des} and the use of different form factors\cite{cheng}
and the possibility of large final state interaction phase\cite{hou}. It may also be 
interesting to consider the new physics effects in those decay modes\cite{neubert}.

  Note that the theoretical description on nonleptonic $B$ decays is model dependent. 
Although the short-distance effects are calculable from  the Wilson coefficients, 
one has to assume factorization approach and adopt some models in evaluating the long-distance
effects. It may then concern many phenomenological parameters, such as the decay constant of $B$ meson,
the transition form factors as well as the effective color number $N_C$, which still suffer from 
large uncertainties. Thus the precision of theoretical calculations is unfortunately limited.

In this paper, we shall consider some less model dependent ways to extract both the weak phase $\gamma$
and the strong phase $\delta$ due to final state interactions. 
The basic point is to assume approximate relations among the strong phases 
and choose the decay modes with relative small EWP effects so that one could ignore their contributions 
as the first step approximation. For this purpose, we take the following four interesting decay modes:  
$B\rarrow \pi^-\pi^+, \pi^0 \pi^+ , \pi^- K^+, \pi^- \bar{K}^0$. It will be seen that, under the above 
assumptions and considerations, the four CP-averaged branching ratios could be used to extract 
the phases $\gamma$ and $\delta$ as well as the relative contributions between strong penguin (SP) graphs 
and tree graphs without additional theoretical inputs. Though such a treatment still suffers from 
some uncertainties, it could directly provide us useful constraints and insight on the phases $\gamma$ 
and $\delta$. We will show that at the 1$\sigma$ level of the current experimental data, 
there exist two correlated regions between $\gamma$ and $\delta$, which are corresponding to two 
solutions of negative $\cos\gamma \cos\delta$, i.e., one solution is with positive $\cos\gamma$ 
but negative $\cos\delta$, another with negative $\cos\gamma$ but positive $\cos\delta$. While 
at more than 1$\sigma$ level a much 
larger region for the phases $\gamma$ and $\delta$ is allowed. 

 Generally, the $B$ decay amplitude can be decomposed by several
$SU(3)$ invariant Fenyman diagrams\cite{zepenfield,wise}. In  this decomposition 
one may see that the amplitudes of decay $B\rarrow \pi\pi$ and $B\rarrow \pi K$ are
correlated. This can be used to study the penguin effectcs as well as the strong 
phases in those modes\cite{sw,xing}.
In $SU(3)$ limits, the decay amplitudes of $B\rarrow \pi^-\pi^+, \pi^0 \pi^+ , \pi^- K^+, 
\pi^- \bar{K}^0$ have  the following forms\cite{gronau,gronau2}

\beqa
{\cal A}(B\rarrow \pi^0\pi^+)&=&\frac{G_F}{\sqrt{2}} V_{ud} V^*_{ub}
(- \frac{T+C}{\sqrt{2}} ) \label{aa} \\
{\cal A}(B\rarrow \pi^-\pi^+)&=&\frac{G_F}{\sqrt{2}} [ V_{ud} V^*_{ub}(T-P)-V_{cd}V^*_{cb}P ]
 \label{pi+-}\\
{\cal A}(B\rarrow \pi^- \bar{K}^0 )&=&\frac{G_F}{\sqrt{2}} V_{ts}V^*_{tb} P' \\
{\cal A}(B\rarrow \pi^- K^+ )&=&\frac{G_F}{\sqrt{2}} [ V_{us}V^*_{ub} T' + V_{ts}V^*_{tb} P']
\eeqa

where the factor $1/\sqrt{2}$ in Eq.(\ref{aa}) comes from the $\pi^0$ wave function.
$T,T'(C)$ and $P,P'$ denote the Tree(Color suppressed) and QCD penguin amplitude with different strong phases:
\beqa
T=|T|e^{i \delta_T}     &,   \ \   P=|P|^{i \delta_P} \nomb \\
T'=|T'|e^{i \delta_T'}  &,    \ \  P'=|P'|^{i \delta_P'}
\eeqa

In the expression for the amplitude ${\cal A}(B\rarrow \pi^-\pi^+)$ in Eq.(\ref{pi+-}), 
the unitarity relation of CKM matrix elements $V_{ud}V^*_{ub}+V_{cd}V^*_{cb}+V_{td}V^*_{tb}=0$ 
has been used to remove the factor $V_{td}V^*_{tb}$ which comes from the inner t-quark of 
the QCD penguins. This allows us to extract the weak phase $\gamma$ instead of $\alpha$, which is 
different from the usual treatments\cite{alpha,wu}. 

The charge conjugate decay amplitude can be obtained by simply inverting the sign of weak phase
$\gamma$. We then get the CP- averaged branching ratios:

\beqa
Br(B\rarrow \pi^+ K^0) &=&\frac{1}{2}
                          \left( B^0\rarrow \pi^+K^0+\bar{B^0}\rarrow \pi^-\bar{K}^0\right)
                          \simeq |V_{ts} V^*_{tb}|^2 |P'|^2\  , \\
Br(B\rarrow \pi^+\pi^0) &=&\frac{1}{2}
                           \left(B^0\rarrow \pi^+\pi^0+\bar{B^0}\rarrow \pi^-\pi^0\right)
                           \simeq \frac{1}{2}|V_{ud} V^*_{ub}|^2 |T+C|^2 \  , \\
Br(B\rarrow \pi^- K^+) &= & 
 \frac{1}{2} \left(Br(B^0\rarrow K^+\pi^-)+Br(\bar{B}^0\rarrow K^-\pi^+)\right)  \nomb\\
&\simeq&|V_{us}V^*_{ub}|^2 \abs{T'}^2 -2 |V_{us}V^*_{ub}||V_{ts}V^*_{tb}| \abs{T'P'} \cos\delta\cos\gamma+|V_{ts}V^*_{tb}|^2 \abs{P'}^2 \ ,\\
Br(B\rarrow \pi^-\pi^+) &= &
\frac{1}{2} \left(Br(B^0\rarrow \pi^+\pi^-)+Br(\bar{B}^0\rarrow \pi^+\pi^-)\right)  \nomb\\
&\simeq&|V_{ud}V^*_{ub}|^2 |(\abs{T} e^{i\delta}-\abs{P})|^2 \nomb\\
& & + 2 |V_{ud}V^*_{ub}||V_{cd}V^*_{cb}| \abs{TP} (\cos\delta\cos\gamma-\pt\cos\gamma) \nomb\\
& & + |V_{cd}V^*_{cb}|^2 \abs{P}^2 \  .
 \eeqa

In writing down the above equations, we have neglected the EWP effects. 
The EWP effects are often thought to be very important\cite{he} in the $B\rarrow \pi K$ decays, 
but it remains depending on different decay modes.  In deed, the EWP effects are of crucial importance in the decay modes: 
$B\rarrow \pi^0 K^0$ and $\pi^0\pi^0$ in which the contributions from the tree diagrams 
are color suppressed. While in the decay modes: $B\rarrow \pi^-\pi^+, \pi^0 \pi^+ , \pi^- K^+, 
\pi^- \bar{K}^0$, the EWP effects are relatively small as the contributions from tree diagrams 
are not color suppressed. As there remain large errors in the current 
experimental data, for simplicity, we may ignore the EWP effects in those four decay modes as 
a good approximation in comparison with the experimental uncertainties. 
To have a quantitative estimation of how good of the approximation, it may be seen from the 
model dependent calculations\cite{kramer}, where the contributions from the EWP graphs were found
to be about $1\%,5\%,5\%, 8\%$ in the decay modes $B\rarrow \pi^-\pi^+, \pi^0 \pi^+ , \pi^- K^+, 
\pi^- \bar{K}^0$, respectively. It is not difficult to recognize that the relative contributions of the 
EWP to SP graphs is about $8\%$, the relative contributions of the tree diagrams to the SP graphs 
is about $40\%$ in the $B\rarrow \pi^- K^+$ decay and is dominated in the $B\rarrow \pi^-\pi^+$ decay.

  It is useful to consider the ratios of the decay rates. Let us define

\beqa
R_1\equiv\frac{Br(B\rarrow \pi^- \bar{K}^0) }{Br(B\rarrow \pi^0 \pi^+)} 
&>& 1.52 \ \ (3.25\pm 1.94) \label{R1} \\
R_2\equiv\frac{Br(B\rarrow \pi^- \bar{K}^0)}{Br(B\rarrow \pi^- K^+)}&=& 1.06 \pm 0.32 \label{R2}\\ 
R_3\equiv\frac{Br(B\rarrow \pi^-\pi^+)}{Br(B\rarrow \pi^- K^+)}&=&0.25\pm 0.1 \label{R3}
\eeqa 

In a naive estimation, the ratio between color suppressed diagram and the 
tree diagram, i.e.  $|C/T|$ is of the order
 ${\cal O}(0.3)$ from the color suppression. However, the model dependent calculation 
show a very small value:$|C/T|\simeq a_2/a_1\simeq 0.05$ when $N_C$ is near 3\cite{kramer}.
By adopting the recent analysis from Ref.\cite{beneke} which is based on the heavy quark
limit, we have $|C/T|\simeq 0.08$.   

In a good approximation, (i.e. neglecting the  terms proportional to 
$|V_{us}V^*_{ub} T/(V_{ts}V^*_{tb} P)|^2 \approx {\cal O}(10^{-2})$ in $\pi K$ modes.)
 $\cos\gamma\cos\delta$ can be solved from the definitions of $R_2$ :
\beq
\cos\gamma\cos\delta\simeq \frac{1}{2} \left|\frac{V_{ts}V^*_{tb}}{V_{us}V^*_{ub}}\right| \ptt 
\left( 1-\frac{1}{R_2}\right)
\label{r2}
\eeq

On with including the leading $SU(3)$ breaking factor $f_\pi/f_K$ in the sense 
of generalized factorization, one then has
\beq
\frac{|P|}{|T|} =\frac{|P'|}{|T'|}, \qquad \frac{|T|}{|T'|} =\frac{f_{\pi}}{f_{K}} 
\eeq
and 
\beq
\delta_T=\delta_{T'} \ , \  \delta_P=\delta_{P'}
\eeq 

Under this approximation, it is then easily seen that the ratio $|P/T|$ can be estimated from 
 $R_1$\cite{gronau,cleoT}:
\beq
\pt\simeq 1.09\times \frac{f_{\pi}}{f_K} \sqrt{\frac{R_1}{2}}
\frac{|V_{ud}V^*_{ub}|}{|V_{ts}V^*_{tb}|}  > 0.055
\label{PT}
\eeq
when taking the central value for the mode $Br(B\rarrow \pi^0 \pi^+) = 5.6$, we have 
$|P/T|=0.08$. The value of  $\abs{P/T}$ can
 also be evaluated from the effective Hamitonian and be 
simply given only by the short distance Wilson coefficients\cite{he,wu} once adopting the 
factorization approach for the hadronic matrix elements
\beq												
{P\over T}=\frac{1}{a_1}\left[ a_4+a_{10}+(a_6+a_8) \frac{2 m^2_{\pi}}{(m_b-m_u)(m_u+m_d)} \right]
\label{pt3}
\eeq
which is found to be 0.05 for $N_C$=3 and $m_u+m_d=1.5$ MeV. Since the validity of Eq.(\ref{pt3})
 only depends on the assumption of factorization, the ratio
$\abs{P/T}$ extracted in this way is helpful to examine how goodness of the factorization approach.
 It seems that the current experimental data prefer a larger $\abs{P/T}$. 
 This needs to be further confirmed by future experiments. 

To naively see the changes of the sign of $\cos\gamma\cos\delta$ as $R_3$ and $R_1$, one may
neglect the terms of the order ${\cal O}(|P/T|^2)$ in $\pi\pi$ decay modes and use the 
modified $SU(3)$ relations. Then $\cos\gamma\cos\delta$ can 
 be simply given in terms of $R_1$ and $R_3$
\beqa
\cos\gamma\cos\delta &\simeq&
1.09 \times\left( \frac{\sqrt{2 R_1}}{4} \right) 
\frac{R_3-1.68/R_1}
     {\frac{|V_{cd}V^*_{cb}|-|V_{ud}V^*_{ub}|}{|V_{ts}V_{tb}|}
      +\frac{|V_{us}V^*_{ub}|}{|V_{ud}V^*_{ub}|}\frac{f_K}{f_{\pi}} R_3
}
\nomb\\
&\simeq &
1.09\times \left(\frac{\sqrt{2 R_1}}{4\lambda}\right)
\frac{R_3-1.68/R_1}{\frac{f_K}{f_{\pi}} R_3+\lambda- 
\left|\frac{V_{ub}}{V_{cb}}\right|}
\label{RR3}
\eeqa
where $\lambda=0.22$  is the Wolfenstein parameter.
This shows  that  $\cos\gamma\cos\delta$ will
change sign when $R_3$ and $R_1$ satisfy the approximate relation $R_3\simeq 1.68/R_1$. 
The precise numerical values of $R_3$ and $R_1$ for  changing the sign of $\cos\gamma\cos\delta$
may be seen from Fig.$\ref{gd2D.ps}$. The values of $R_3$ is slightly higher than  the ones 
 by Eq.(\ref{RR3}).   

With the above considerations, the phases $\gamma$ and $\delta$ can be extracted 
from $R_1$, $R_2$ and $R_3$ . As the equations
are quadratic in $\cos\gamma$ and $\cos\delta$, there exists a twofold ambiguity in determining
these two phases.
In Fig.$\ref{gd2D.ps}$, we present a contour plot for $R_2$ and $R_3$ in the 
$\cos\gamma-\cos\delta$ plane with $R_1$ being fixed at 3.0, 4.5, 6.0, 7.5. 
Where the solid and dashed contours
correspond to different values of $R_2$ and $R_3$. The points at which the two kind of curves intersect
are the solutions of $\cos\gamma$ and $\cos\delta$. It can be seen from  Fig.1 that
these contours change largely for different values of $R_2$ and $R_3$.
When $R_2<1 $ the contours of $R_2$ and $R_3$ are all in the I$\!$I and I$\!$V quadrants. 
When $R_2>1 $ the contours move into the I and I$\!$I$\!$I quadrants. This behavior can be
understood  from Eq.(\ref{r2}).
Thus $\cos\gamma \cos\delta$ will change sign when $R_2$ moves across the point $R_2 = 1$. The changes of
$R_3$ contours also have the similar reason. From the present data within 1$\sigma$ 
standard deviation, $R_2$ and $R_3$ are in the range $0.74\alt R_2\alt 1.38$ and
 $0.15\alt R_3 \alt 0.35 $, respectively
. Since $R_3$ is smaller than 0.35 at $1\sigma$ level, $\cos\gamma\cos\delta$ will 
be negative for small $R_1$. Namely a negative $\cos\gamma$ corresponds to a strong phase $\delta$ 
in the first quadrant, for positive $\cos\gamma$, the angle $\delta$ becomes large and takes values 
in the second quadrant. From Fig.1,  one may see that for large $R_1$, a positive solution of 
$\cos\gamma\cos\delta$ is also allowed. For $R_1\agt 0.75$, the allowed
range of $\cos\gamma$ and $\cos\delta$ becomes large and lies in the region: $0.2\alt \cos\gamma \alt 1$ and
 $ -1 \alt \cos\delta \alt 1$
or $-1 \alt \cos\gamma \alt 0.1 $ and $ -1\alt \cos\delta \alt 1 $.  

The constraints on the phase $\gamma$ may also come from other experiments, 
such as $B^0_{d,s}-\bar{B}^0_{d,s}$ mixing, CP-violating parameter $\epsilon_K$ in the kaon decay, 
and CKM matrix element $V_{ub}$ from semileptonic $b\rarrow u$ decays. Combining all the constraints
together and taking the branching ratio for the $B\rarrow \pi^+\pi^0$ decay to be
$Br(B\rarrow \pi^+\pi^0)=5.6^{+2.6}_{-2.3}\pm 1.7$, the allowed region for $\gamma$ is 
shown in Fig.\ref{rhoeta.ps}. 

 It is found that the allowed range for $\gamma$ is: 
$35^{\circ}\alt  \gamma \alt 62^{\circ}$ or $86^{\circ} \alt \gamma \alt 151^{\circ}$,
the corresponding values for the phase $\delta$ could range from 
$106^{\circ} $ to $180^{\circ}$ or from $0^{\circ}$ to $75^{\circ}$. The allowed regions for 
the phases $\gamma$ and $\delta$ are plotted in Fig. 3 and given by the two shadowed ones. 
One can see from the figure that large region for $\cos\gamma$ and $\cos\delta$ 
has been exluded from $R_1$, $R_2$ and $R_3$ when they are at 1$\sigma$ level.

 Recently, CLEO collaboration has also reported the data on direct CP violation in 
$B\rightarrow \pi^- K^+$ decay. Let us now consider CP asymmetries in both $B\rarrow \pi\pi$ and $B\rarrow 
\pi K$ decays. They are defined as
\beqa
a_{\epsilon''}^{(\pi^- K^+)} &=& \frac{\Gamma(\bar{B}^0\rarrow  \pi^-K^+)
                                       -\Gamma(B^0\rarrow \pi^+K^-)} 
                             {\Gamma(\bar{B}^0\rarrow \pi^-K^+)+\Gamma(B^0\rarrow \pi^+K^-)} \nomb \\                         
                    &=& \left(2 |V_{us}V^*_{ub}V_{ts}V^*_{tb}|\pt \sin\gamma\sin\delta \right)\nomb \\
                    & &   \times \left(|V_{us}V^*_{ub}|^2-2|V_{us}V^*_{ub}V_{ts}V^*_{tb}|\pt 
                          \cos\gamma\cos\delta +|V_{ts}V^*_{tb}|^2 \pt^2 \right)^{-1},\\
a_{\epsilon''}^{(\pi^+\pi^-)} &=& \frac{\Gamma(\bar{B}^0\rarrow \pi^+\pi^-)-\Gamma(B^0\rarrow \pi^+\pi^-)}
                             {\Gamma(\bar{B}^0\rarrow \pi^+\pi^-)+\Gamma(B^0\rarrow \pi^+\pi)} \nomb\\
                    &=& \left(2 |V_{ud}V^*_{ub}V_{cd}V^*_{cb}| \pt \sin\gamma\sin\delta\right)\nomb \\
                    & & \times \left(|V_{ud}V^*_{ub}|^2 (1-2 \pt\cos\delta+\pt^2)
                       +|V_{cd}V^*_{cb}|^2 \pt^2 \right.\nomb\\
                    & &\left. +2|V_{ud}V^*_{ub}V_{cd}V^*_{cb}| \pt\cos\gamma
                       (\cos\delta-\pt)\right)^{-1}
\eeqa
Here we have used the notation for the general rephase-invariant CP-violating observables 
classified in \cite{PW,wu-cp}. As $|P/T|$ is at order of ${\cal O}(10^{-1})$, for an approximate 
estimation, one may neglect the $|P/T|$ terms in the denominator, thus the above formulae
are simplified 
\beqa
a_{\epsilon''}^{(\pi^- K^+)} &\approx& 2 \frac{|V_{us}V^*_{ub}|}{|V_{ts}V^*_{tb}|} 
\left| \frac{T}{P}\right| \sin\gamma\sin\delta \\
a_{\epsilon''}^{(\pi^+\pi^-)} &\approx& 2 \frac{|V_{ud}V^*_{ub}|}{|V_{cd}V^*_{cb}|} \pt 
\sin\gamma\sin\delta \nomb\\
&\approx& 0.59 \times \frac{f_{\pi}^2}{f_K^2} R_1 a_{\epsilon''}^{(\pi^- K^+)} 
\label{acp-pi}
\eeqa
which implies that  
$a_{\epsilon''}^{(\pi^+\pi^-)}$ may become large with $R_1$ increasing.
From the data reported by the CLEO collaboration, no significant deviation from zero was observed:
$a_{\epsilon''}^{(\pi^- K^+)}= -0.04\pm 0.16$\cite{cleoCP}.
Even at $90\%$ CL, $a_{\epsilon''}^{(\pi^- K^+)}$ is limited in the range [-0.30,0.22].
Incoporating this result, the allowed regions for the phases $\gamma$ and $\delta$ are further constrained
, which is shown in Fig. 3.  It is seen that some regions have further been excluded when
$a_{\epsilon''}^{(\pi^- K^+)}$ has the value within the 1$\sigma$ standard deviation.
If the values of $R_1$, $R_2$ and $R_3$ are taken to be at 2$\sigma$ level, 
$ a_{\epsilon''}^{(\pi^-K^+)}$ could be as large as the experimental bound given 
at 90$\%$ CL.  From Eq.($\ref{acp-pi}$) and $(\ref{PT})$, 
the maximum value of $a_{\epsilon''}^{(\pi^+\pi^-)}$ is approximately given by
\beqa
a_{\epsilon''}^{(\pi^+\pi^-)}|_{max}&\simeq& 
0.59 \frac{f_{\pi}^2}{f_K^2} R_1^{(max)} \times a_{\epsilon''}^{(\pi^- K^+)}|_{max} \nomb\\
&\alt& 0.40, \quad \mbox{ (at 1$\sigma$ level)  }
\eeqa
Where $R_1$ is taken to be within the 1$\sigma$ standard deviation. 
The numerical results for $a_{\epsilon''}^{(\pi^+\pi^-)}$
and $a_{\epsilon''}^{(\pi^- K^+)}$  are plotted in Fig.\ref{Acp2D.ps} 
and Fig.{\ref{Acp2Dk.ps} as functions of the ratios $R_1$ and $R_3$. 
It can be seen that for  $R_1\agt 2.65$ one has $|a_{\epsilon''}^{(\pi^+\pi^-)}|
> |a_{\epsilon''}^{(\pi^- K^+)}|$.

In conclusion, assuming $SU(3)$ symmetry for the strong phases and ignoring 
the relative small EWP effects in the $B\rarrow \pi^-\pi^+, \pi^0 \pi^+ , \pi^- K^+, 
\pi^- \bar{K}^0$ decays, a model independent approach is proposed to extract the weak phase  
$\gamma$ and the strong final interacting phase $\delta$.
From  the present data a negative $\cos\gamma\cos\delta$ is favored. Two solutions for the 
phases $\gamma$ and $\delta$ have been obtained at 1$\sigma$ level of the current experimental data, 
though their allowed regions have been strongly restricted, there remain large uncertainties, 
two interesting allowed regions for the phases $\gamma$ and $\delta$ have been obtained at 
the 1$\sigma$ level. The numerical values of the phases $\gamma$ and $\delta$ have been found to lie in the
regions: $35^{\circ}\alt\gamma\alt 62^{\circ}$ and $106^{\circ}\alt \delta \alt 180^{\circ}$ or 
 $86^{\circ}\alt\gamma\alt 151^{\circ}$ and $0^{\circ}\alt\delta\alt 75^{\circ} $ 
We would like to point out that with large uncertainties of the current experimental data at 
more than 1$\sigma$ level, one cannot exclude solutions with a small strong phase $\delta$ and 
the values of $\gamma$ constrained from $V_{ub}$, $B^{0}_{d,s}-\bar{B}^{0}_{d,s}$ mixings and 
indirect CP-violating parameter $\epsilon_K$ within the standard model. 
The direct CP asymmetries $ a_{\epsilon''}^{(\pi^+\pi^-)}$ and $a_{\epsilon''}^{(\pi^- K^+)}$
in $B\rarrow \pi^+\pi^-$ and $\pi^-K^+$ decays have also be estimated. within the errors of
the measurement of $a_{\epsilon''}^{(\pi^- K^+)}$, the maximum value of 
$ a_{\epsilon''}^{(\pi^+\pi^-)}$ could be as large as 40$\%$, a larger value may be
possible if $R_1$ is large. The  more precise experimental data in the $B\rarrow \pi \pi$ 
and $\pi K$ decays will be very plausible for extracting the important weak phase $\gamma$ and 
strong phase $\delta$ as well as testing how good of the factorization approach. It may also 
provide us a possible window for new physics with new CP-violating sources\cite{ww} 
which could change all the constraints arising from the $B^{0}_{d,s}-\bar{B}^{0}_{d,s}$ 
mixings, radiative rare $B$ decays ($b\rarrow s\  \gamma$) and observed CP-violating 
parameters in the kaon decays as well as from the decay amplitudes of hadrons\cite{wz}.
Finally, we would like to address that our current results have been obtained by   
assuming the SU(3) relations among the strong phases and ignoring the EWP effects in 
the considered four decay modes. To precisely extract the phases $\gamma$ and $\delta$, 
one needs to improve not only the experimental measurements but also theoretical 
approaches which is going to be investigated elsewhere. 

{\bf Acknowledgments }
We would like to thank David E. Jaffe at CLEO for very helpful comments concerning the data for
the direct CP violation. This work was supported in part by the NSF of China under grant No. 19625514.

\begin{figure}
\centerline{ \psfig{figure=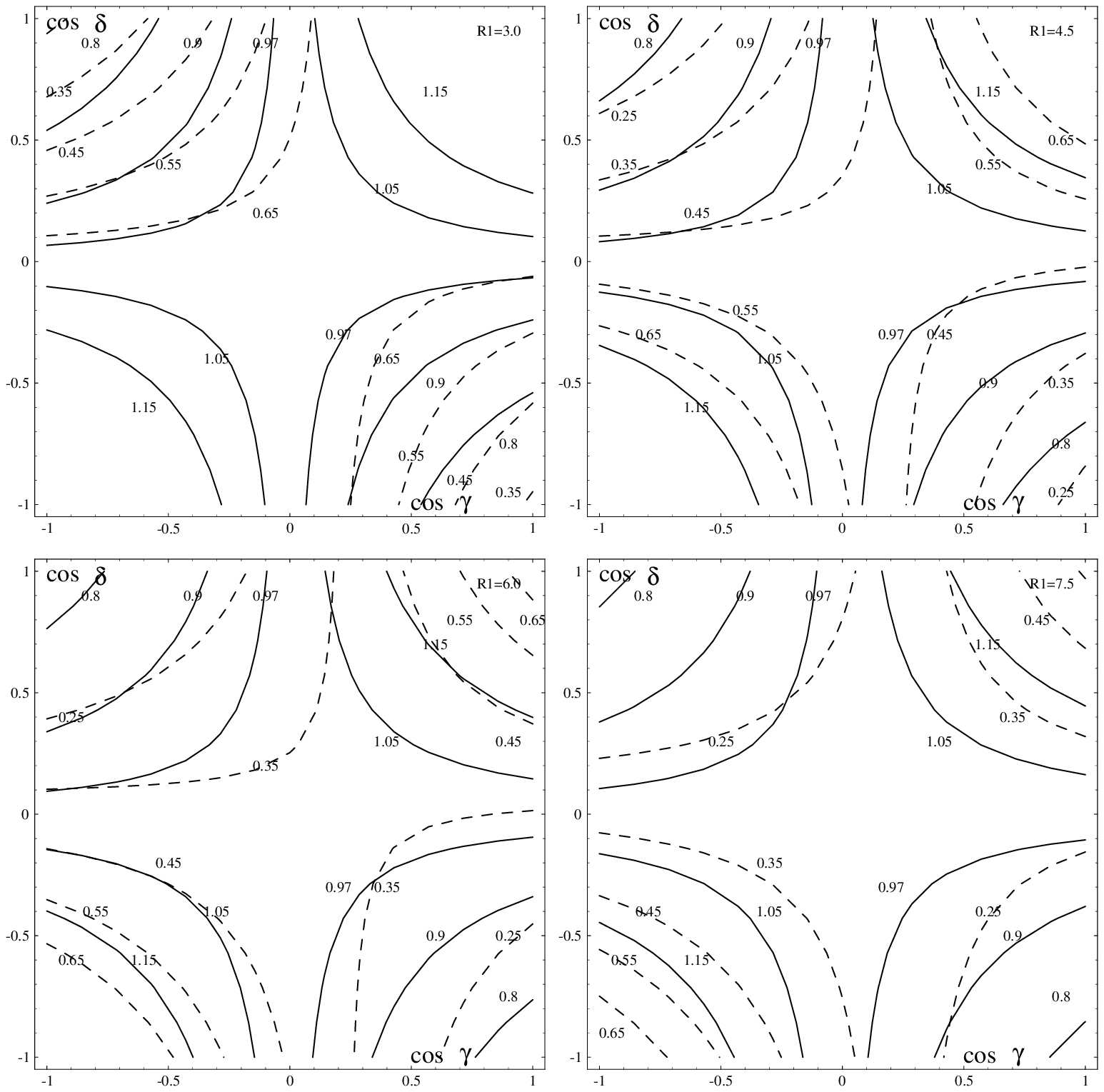, width=14 cm} }
 \caption{
The contours of $R_2$(solid) and $R_3$(dashed) in $\cos\gamma-
\cos\delta$ plane. The five solid(dashed) curves correspond to $R_2$=0.8,
 0.9, 0.97, 1.05, 1.15($R_3$= 0.25, 0.35, 0.45, 0.55, 0.65) respectively with $R_1$
varies from 3.0, 4.5, 6.0 to 7.5.
}
\label{gd2D.ps}	
 \end{figure}

\begin{figure}
 \centerline{ \psfig{figure=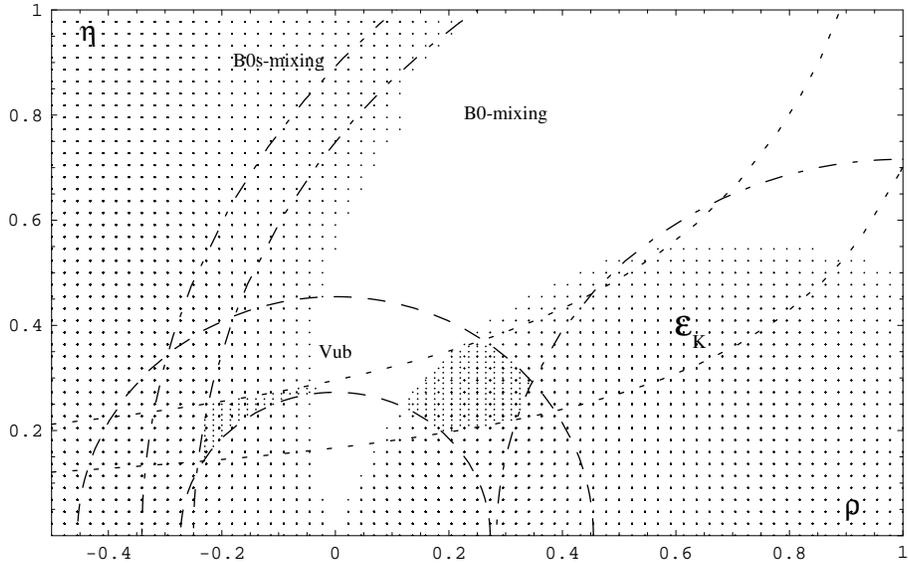, width=12 cm} }
 \caption{ 
The allowed region in $\rho-\eta$ plane. The shadowed area corresponds to
the allowed region from the constraints of $R_1,R_2$ and $R_3$ at 1 $\sigma$ level. Other constraints
are from $B^0_{d,s}$,$\epsilon_K$ and $V_{ub}$. The dark area is the allowed region
with all the constraints included.
}
\label{rhoeta.ps}	
 \end{figure}

\begin{figure}
 \centerline{ \psfig{figure=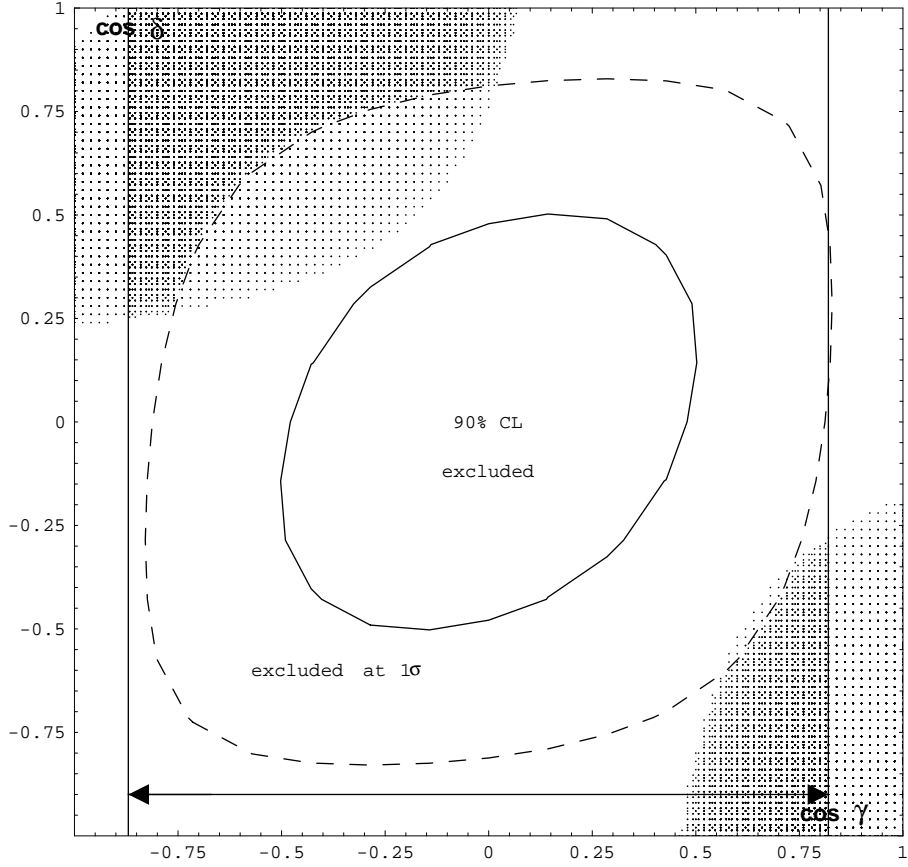, width=12 cm} }
 \caption{
The allowed regions of the phases $\gamma$ and $\delta$ in $\cos\delta-\cos\gamma$ plane. 
The whole shadowed areas are the allowed region from the constraints of 
the ratios $R_1$, $R_2$ and $R_3$  at 1$\sigma$ level with  $|V_{ub}/V_{cb}|=0.08\pm 0.02$. 
The regions within the closed curves are corresponding to the ones exluded by the data of 
$a_{\epsilon''}^{(\pi^- K^+)}$ at 1$\sigma$ level (dashed one) and at 90$\%$ CL (solid one), respectively. 
The range within the two vertical lines is the allowed range for the phase $\gamma$ constrained 
from $B^0_{d,s}-\bar{B}^0_{d,s}$ mixings, $\epsilon_K$ and $V_{ub}$. The dark shadowed regions are
the allowed ones for the phases $\gamma$ and $\delta$ from the whole constraints.
\label{gd.ps}
}
\label{gd2d.ps}	
 \end{figure}

\begin{figure}
 \centerline{ \psfig{figure=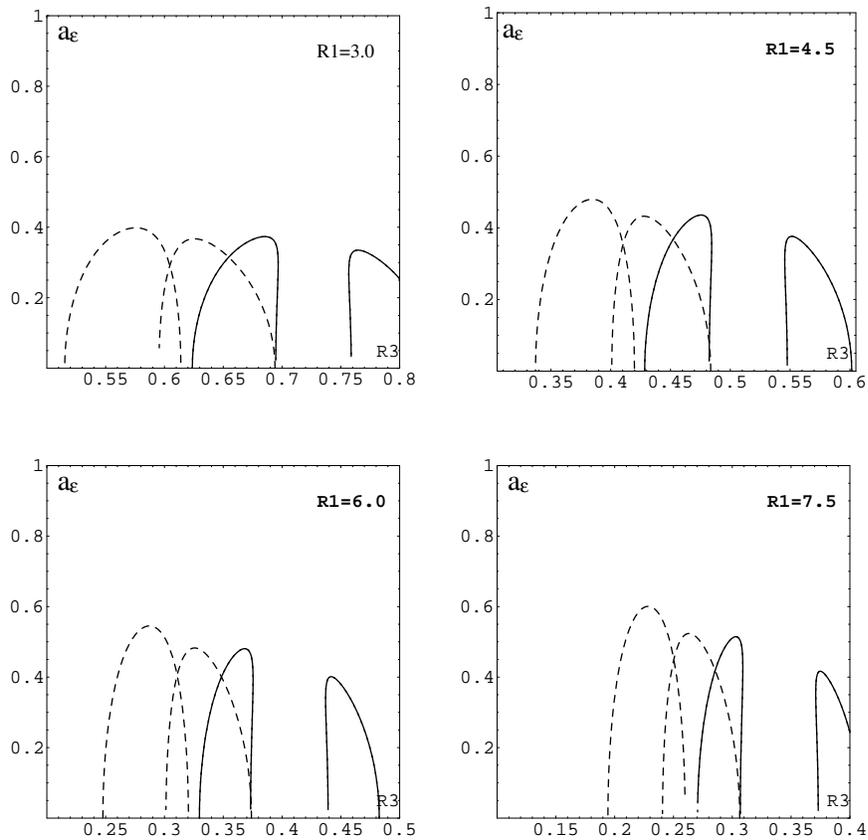, width=12 cm} }
 \caption{
 $ a_{\epsilon''}^{(\pi^+\pi^-)}$ vs $R_3$ with different values of $R_1$. The
dashed and  solid  curves correspond to $R_2$=0.95 and 1.05 respectively 
}
\label{Acp2D.ps}	
 \end{figure}
\begin{figure}
 \centerline{ \psfig{figure=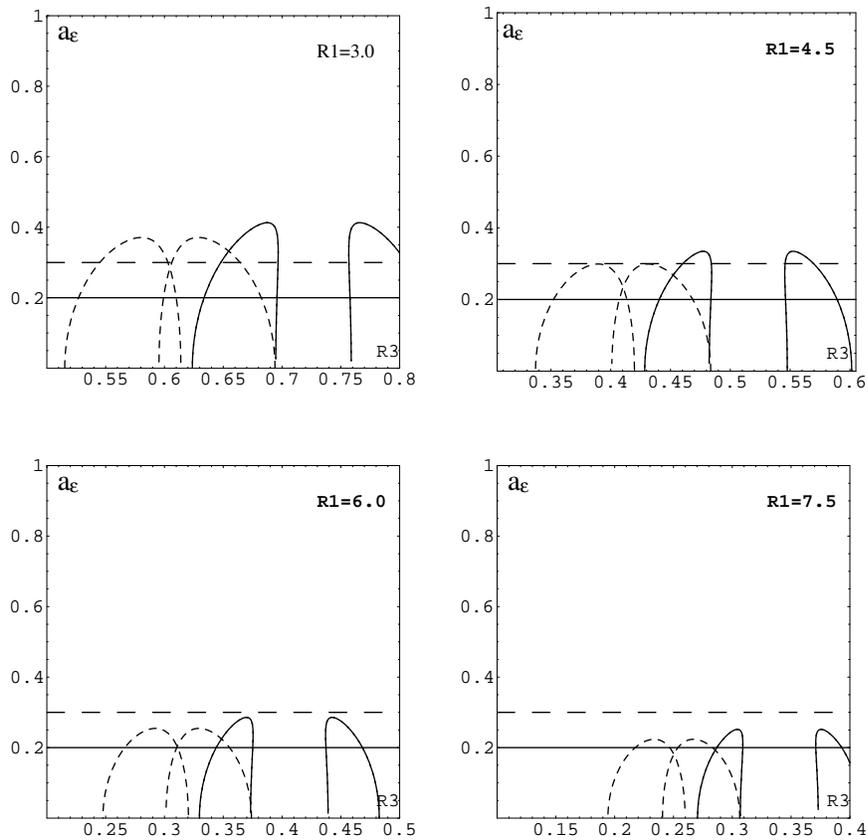, width=12 cm} }
 \caption{
The same as in Fig.\ref{Acp2D.ps}, but for $ |a_{\epsilon''}^{(\pi^- K^+)}|$.
The two horizontal lines indicate the upper bound from the 
CLEO data at 1$\sigma$(solid)
and 90$\%$ CL respectively.
 }
\label{Acp2Dk.ps}	
 \end{figure}
\end{document}